# Study on the multiple characteristics of $M_3$ generation of pea mutants obtained by neutron irradiation

Dapeng Xu[1,3*], Ze'en Yao[1,3*], Jianbin Pan[2], Huyuan Feng[2], Zhiqi Guo[1,3], Xiaolong Lu[1,3]

[1]School of Nuclear Science and Technology, Lanzhou University, Lanzhou 730000, China; [2]School of Life Sciences, Lanzhou University, Lanzhou 730000, China; [3]Engineering Research Center for Neutron Application Technology, Ministry of Education, Lanzhou University, Lanzhou 730000, China

[*]Corresponding author: Ze'en Yao (email: zeyao@lzu.edu.cn), Dapeng Xu (email: xudp@lzu.edu.cn)

**Abstract:** Irradiation breeding is an important technique in the effort to solve food shortages and improve the quality of agricultural products. In this study, a field test was implemented on the $M_3$ generation of two mutant pea plants gained from previous neutron radiation of pea seeds. The relationship between agronomic characteristics and yields of the mutants was investigated. Moreover, differences in physiological and biochemical properties and seed nutrients were analyzed. The results demonstrated that the plant height, effective pods per plant, and yield per plant of mutant Leaf-M1 were 45.0%, 43.2%, and 50.9% higher than those of the control group. Further analysis attributed the increase in yield per plant to the increased branching number. The yield per plant of mutant Leaf-M2 was 7.8% higher than that of the control group, which could be related with the increased chlorophyll content in the leaves. There was a significant difference between the two mutants in the increase of yield per plant owing to morphological variation between the two mutants. There were significant differences in SOD activity and MDA content between the two mutants and the control, indicating that the physiological regulation of the two mutants also changed. In addition, the iron element content of seeds of the two mutants were about 10.9% lower than in the seeds of the control group, a significant difference. These findings indicate that the mutants Leaf-M1 and Leaf-M2 have breeding value and material value for molecular biological studies.
**Keywords** neutron radiation, irradiation breeding, pea mutant, agronomic characteristic, yield

## 1. Introduction

Food shortages are a worldwide problem. According to the FAO (Food and Agriculture Organization of the United Nations) report on "The State of Food Security and Nutrition in the World (2019)," 820 million people suffer from hunger worldwide [1]. Irradiation breeding is an important solution to food shortages. In 1928, Stadler, an American scientist, carried out a radiation study on corn and wheat and proved for the first time that X-rays could induce mutations, opening a new research field radiation-induced plant mutation [2]. Since then, scientists worldwide have studied radiation mutation breeding and the biological effects caused by radiation. Much research has been conducted in the last decade [3-11]. According to statistical data from the FAO and International Atomic Energy Agency (IAEA), a total of 3,322 plant mutant species have been developed around the world as of March 2020 [12]. The successful application of mutant species can increase crop yields and improve the quality of agricultural products, promoting the development of the agricultural economy.

Neutron irradiation breeding is an important part of the research field of irradiation breeding. Neutron rays are a kind of high LET (Linear Energy Transfer) ray [13]. As compared with X-ray and γ-ray radiation, neutron radiation has the characteristics of strong penetrability, wide mutation spectra, high rate of variation, and relatively stable characteristics of the offspring of variation, producing greater and more obvious biological effects than γ-ray [14-15]. Marshak was the pioneer of neutron radiation of plants or plant cells. In 1939, he used fast neutrons to irradiate root tip cells of tomato plants and found that the population of normal late cells decreased to the lowest point after 3 to 9 h of neutron radiation [16]. Subsequently, international research institutions have studied the biological effects of mutation breeding through neutron irradiation of plants via various techniques. The relevant biological effects have undergone preliminary analysis, and abundant research fruits have been achieved over the decades. Recent research achievements are introduced in the References [17-23].

As one of three major legume crops in the world [24], the pea (*Pisum sativum* L.) plant has a direct impact on human life and productivity. It has been a scientific research hotspot [25-29] since Mendel used it to derive the basic law of genetics [25]. In our previous study on neutron-radiated pea seeds, two mutants with significant morphological changes were obtained which maintained the morphological characteristics of variation throughout their life. In the following study, we will evaluate the breeding values and molecular biology research potential of these two mutants through analysis of their physiological, biochemical, and agronomic characteristics, particularly their yield.

## 2. Materials and Methods

### 2.1 Materials

In 2014 and 2015, we irradiated pea seeds with $^{252}$Cf isotopic neutrons [30]. We used the needle-leafed pea MZ-1 (*Pisum sativum* L. var. MZ-1), a leafless variety imported by the Institute of Soil Fertilizer and Water-saving Agriculture, Gansu Academy of Agricultural Sciences [31]. The absorbed dose of neutron radiation administered to the seeds was 0.51-9.27 Gy. During the $M_1$ Generation field experiment in 2016, we found two mutant plants with significant variation in leaf characteristics through screening and analysis of mutants. As shown in Fig.1, the needle-shaped leaves of one mutant pea changed to broad leaves (here named as Leaf-M1) and the leaves of the other mutant pea changed from dark green to emerald green (here named as Leaf-M2). In a subsequent $M_2$ Generation field experiment, morphological observation and DNA molecular analysis of self-crosses verified that the novel characteristics of the two mutants had high genetic stability.

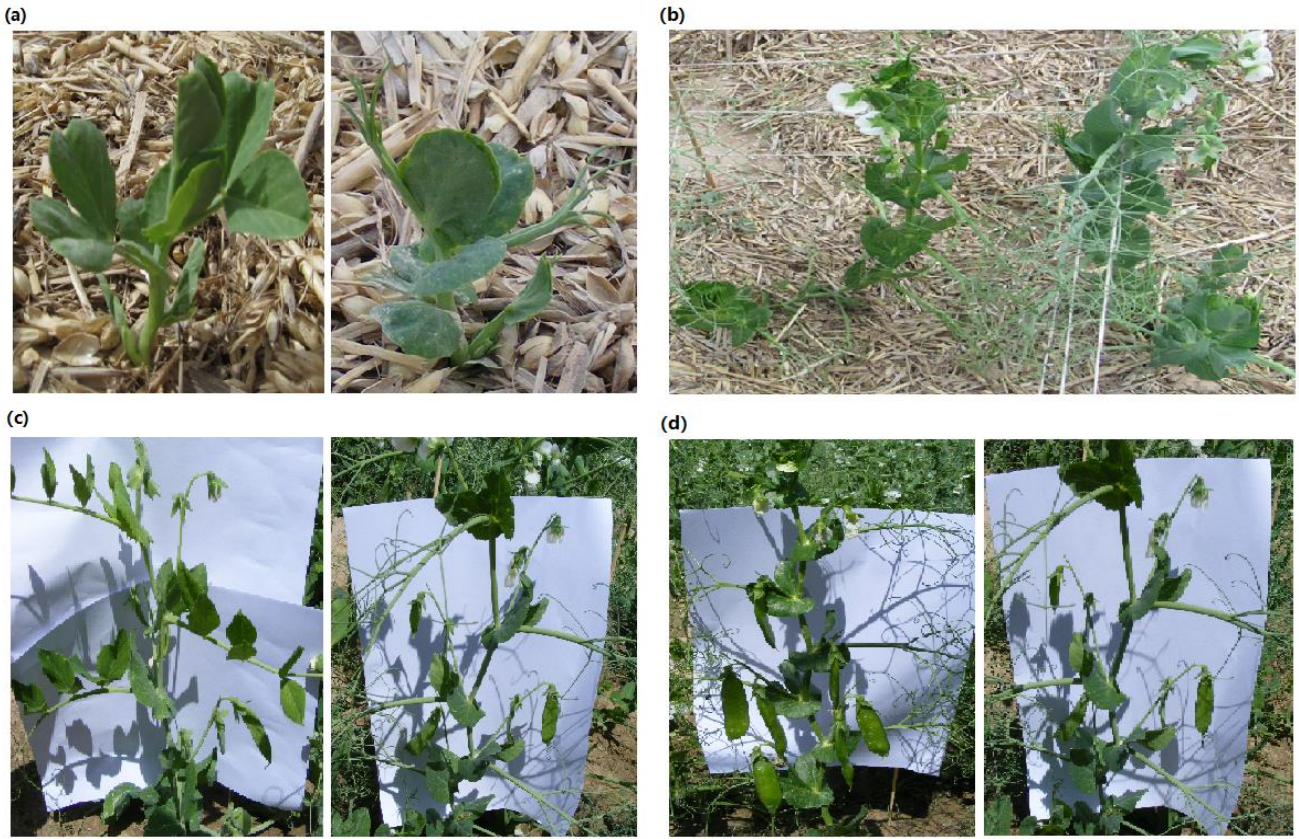

**Fig.1** Two mutant peas and control peas. (a)&(c): The left plant is the mutant Leaf-M1 and the right is the control pea MZ-1; (b)&(d): The left is the mutant Leaf-M2 and the right is the control pea MZ-1.

## 2.2 Field trials and methods

Leaf-M1, Leaf-M2, and MZ-1 seeds were sown in a field in the Yuzhong Campus of Lanzhou University on April 6$^{th}$, 2017, for the field experiment of the M$_3$ generation. Five replications were set up with 20 peas per replicate. Unified operation was adopted for the whole experimental field. Growth conditions during the seedling stage were recorded and the agronomic characteristics of the pea plants (including plant height, branching number, yield per plant, effective pod number per plant, pod number per branch, and yield per pod) were measured at three different stages. The parameters of plant height and yield were measured using a rule with 1 mm to 1000 mm scales and an electronic balance with 0.1 g to 999 g mass scales respectively. The harvested M$_3$ generation pea seeds were also tested for composition. Water content was tested by direct drying method, protein was tested by Kjeldahl method, crude fat was tested by residual method, and crude starch was tested by acid hydrolyzation method. Phosphorus (P) content was tested by acid solution-Vanadium molybdate yellow colorimetry. Potassium (K) content was detected by acid solution-flame photometry. Iron (Fe), Zinc (Zn), Manganese (Mn), Copper (Cu), and Magnesium (Mg) were tested by flame atomic absorption spectrometry. Tests were completed with the assistance of the Agricultural Test Center of Gansu Academy of Agricultural Sciences, from which the test report was issued.

## 2.3 Experimental analysis on physiological and biochemical characteristics

At the late stage of pea seedling formation, a quantity of leaves were cut from each replicate for analysis of the physiological and biochemical characteristics of the peas. The analyzed characteristics included SOD (Superoxide Dismutase) activity, MDA (Malondialdehyde) content, reducing sugar content, and chlorophyll relative content. Details are as follows.

SOD activity in pea leaves was analyzed by NBT method. First, 0.5 g pea leaves were rinsed in distilled water and cut into pieces, then placed in a pre-cooled pot. Next, 5 ml cooled 50 mmol/L phosphate buffer was added in 3 steps to prepare a homogenate. The homogenate was then centrifuged for 10 min on a centrifugal machine at 4000 g. The resulting supernate was the crude enzyme solution, for which the total volume was measured. Four test tubes of supernate were prepared, including two as the experimental group and two as the control group (replacing enzyme solution with buffer). Prepared reaction solution was added to the four test tubes and mixed evenly. One control test tube was kept in darkness, while the remaining 3 test tubes were reacted for 10 min under a fluorescent lamp. At the end of the reaction, the three test tubes were placed in darkness for 5 min to terminate the reaction. The test tube that remained in darkness was used as the blank control. OD values of the other three test tubes at 560 nm were tested to calculate SOD activity.

MDA content in pea leaves was analyzed through TBA method. First, 0.2 g pea leaves were rinsed in distilled water, wiped dry, and placed in a pot. Next, 5 ml distilled water was collected by a measuring cylinder before grinding the pea leaf. About 2 ml distilled water was added to the pot for grinding. The ground mixture was transferred to centrifugal tubes (10 ml range). The remaining 3 ml distilled water in the measuring cylinder was used to rinse the pot three times, and 5 ml 0.5% TBA was added to all centrifugal tubes, for a final volume of 10 ml solution. All samples were treated with a 10 min boiling water bath, followed by 10 min centrifuging at 4000 g after cooling. Absorbance values of the supernate at 450 nm, 532 nm, and 600 nm were measured to calculate MDA content.

Reducing sugar content in leaves was analyzed by DNS method. First, glucose standard solution with a known concentration was collected to react with 3,5-dinitrosalicylic acid reagent. Spectrophotometry was measured to draw a standard curve. The experimental steps of the standard curve are detailed in the Reference [32]. Second, 0.5 g pea leaves and 2 to 3 ml distilled water were mixed in a pot to grind the leaves completely. The pot was rinsed with 2 to 3 ml distilled water three times (overall volume <10 ml) and the material was transferred to test tubes. All test tubes were placed in a 50°C water bath for 20 min, then cooled, and then centrifuged for 5 min at 4000 g. Supernate was collected and transferred to new test tubes, then dissolved to a constant volume of 20 ml in distilled water. These are reducing sugar testing solutions. Next, 2 ml of reducing sugar testing solution was collected and operated according to the methodology for making glucose standard curves. Finally, reducing sugar content was calculated according to the standard curve.

Chlorophyll relative content in the leaves was measured directly with a SPAD-502 Chlorophyll meter.

**2.4 Data analysis methods**

The emergence rate was defined as the percentage of the number of emergent peas divided by the number of seeding peas, and the seedling formation rate was defined as the percentage of the number of peas developing to five pairs of stipules or more divided by the number of seeding peas.

One-way ANOVA and LSD multiple comparisons were conducted in SPSS Statistics 17.0 software (SPSS Inc., Chicago, IL, USA) for analysis of experimental data. Standard errors for all experimental groups were calculated and relevant graphs were drawn by using Excel 2007 software.

**3. Results and discussion**

**3.1 Analysis of seedling and growth characteristics**

Growth of pea plants during the seedling stage was recorded comprehensively. Statistical analysis of emergence rate was conducted every six days from the initial seedling stage, for a total of six statistical analyses. Data processing results are shown in Fig.2a. The emergence rates of the three groups were generally similar in the first statistical analysis. However, the emergence rate of Leaf-M1 was higher than that of the control group at the second statistical analysis and the emergence rate of MZ-1 was higher than that of Leaf-M2. Moreover, there were significant differences between Leaf-M1 and Leaf-M2. This relationship among the three groups was also present in the third statistical analysis, with the exception of the significant difference between Leaf-M1 and Leaf-M2. The relationship among the three groups remained stable at the fourth, fifth, and sixth statistical analyses: the emergence rate of Leaf-M1 was higher than that of Leaf-M2 and the emergence rate of Leaf-M2 was higher than that of MZ-1. As the plants developed, the differences among the three groups gradually decreased, and the emergence rates of three groups were very similar at the sixth analysis. These findings demonstrate that seedling emergence of Leaf-M1 occurs quickly, and the development trend for Leaf-M2 is similar to that of the control. Finally, the emergence rates of three groups effectively converge at the sixth statistical time point.

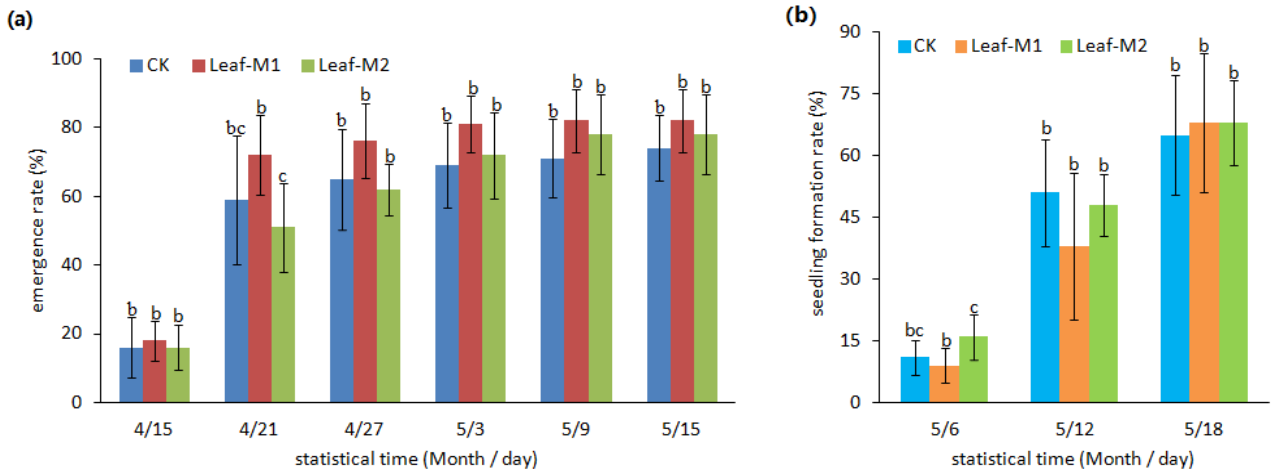

**Fig.2** Emergence rates and seedling rates of two mutant peas and control peas at different statistical analyses. Experiments were performed at least five times. Values are the mean ± SD (n = 5), and bars indicate the SD. Different lowercase letters indicate significant differences ($P < 0.05$)

A statistical analysis of seedling formation rates was conducted every six days from the early days of seedling formation, and three statistical analyses were conducted. Results are shown in Fig.2b. On May 6th, the seedling formation rate of Leaf-M2 was higher than that of the control group and the seedling formation rate of the control group was higher than that of Leaf-M1. In addition, there were significant differences in seedling formation rate between Leaf-M2 and Leaf-M1. The data from May 12th showed that the seedling formation rates of Leaf-M2 and the control group were similar, and both were slightly higher than that of Leaf-M1. By May 18th, the seedling formation rates of the three groups were approximately consistent. Leaf-M1 seedling morphology developed slowly, while the seedling morphology of Leaf-M2 and MZ-1 developed at similar rates. At the end of the trial, the seedling formation rates of the three groups were approximately equivalent.

### 3.2 Comparison of branch development of three groups

Pea plants generally develop a branching habit. Growth time and quantity of branch development affect the relevant biomass of peas, including yields. Branches of all three groups were measured during the late initial seedling stage, growth stage, and harvest stage. Results of statistical analysis are shown in Fig.3a. As shown in that figure, during the late initial seedling stage (sampling date: April 29th), Leaf-M2 showed the greatest branch number per plant (0.82), followed by 0.47 in the control group, with the lowest number of branches per plant (0.13) belonging to Leaf-M1. The differences in branch number per plant were extremely significant between the three groups. At the growth stage (on May 13th), the size relationship among the three groups was concurrent with that measured in the late initial seedling stage, and the branch numbers per plant for Leaf-M2, the control group, and Leaf-M1 were 0.89, 0.76, and 0.32, respectively. However, no significant difference between Leaf-M2 and MZ-1 in terms of branching number per plant was noted. In addition, the branching numbers per plant for Leaf-M2 and the control group were nearly equivalent (2.73 and 2.69, respectively) at harvest time (July 8$^{th}$). The branch number per plant for Leaf-M1 was 4.07, 51.3% and 49.1% higher than those of Leaf-M2 and MZ-1, respectively. According to difference significance analysis, Leaf-M1 showed significant differences with Leaf-M2 and the MZ-1. This finding reflects the fact that Leaf-M2 develops branches earlier, but its branching number per plant became similar with that of MZ-1 in subsequent development. Although the branches of Leaf-M1 developed relatively late, the final branching number per plant was significantly higher than that of MZ-1.

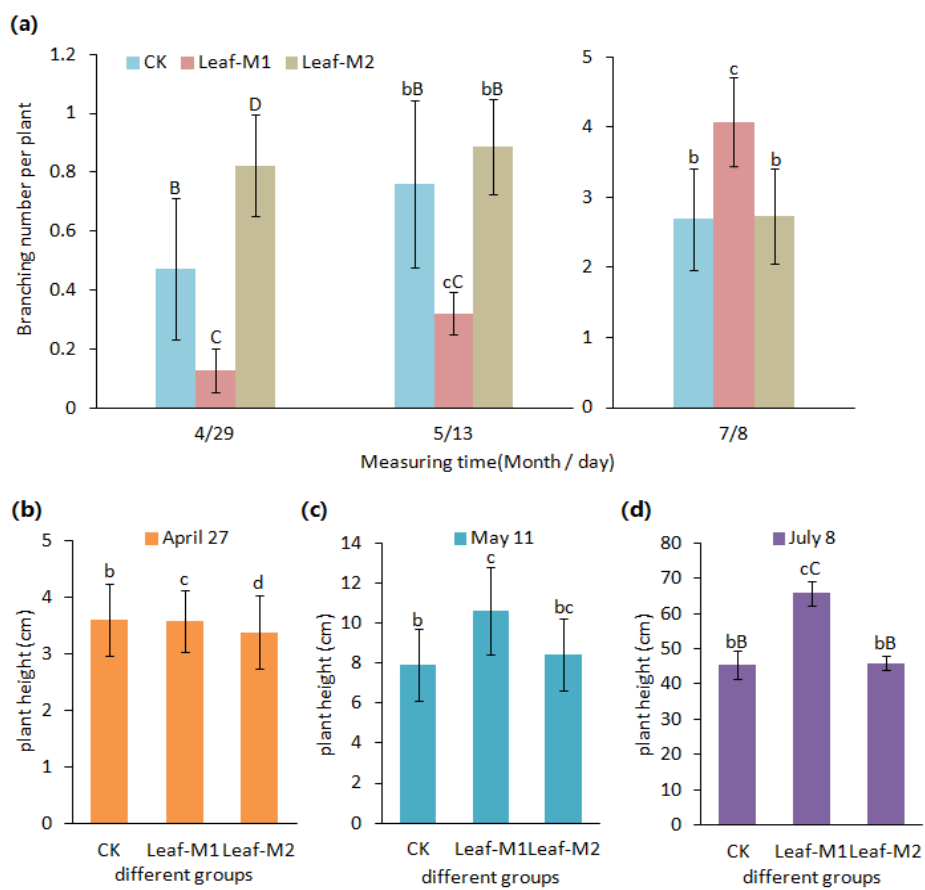

**Fig.3** Branching number per plant and plant heights of two mutants and control peas during different development periods. Experiments were performed at least five times. Values are the mean ± SD (n = 5), and bars indicate the SD. Different capital letters indicate extremely significant differences ($P < 0.01$), and different lowercase letters indicate significant differences ($P < 0.05$)

### 3.3 Comparison of plant height among three groups

Height is an important morphological trait for plants. Plant heights of the three groups were measured on April 27$^{th}$, May 11$^{th}$, and July 8$^{th}$. Results following data processing are shown in Fig.3b,

Fig.3c, and Fig.3d, respectively. As shown in Fig. 3b, on April 27th, the plant height of MZ-1 was the highest, and the plant height of Leaf-M2 was the lowest. However, no significant difference was observed. On May 11th, the plant height of Leaf-M1 became the highest, and it was significantly higher than that of MZ-1. Meanwhile, the plant height of Leaf-M2 was similar to that of MZ-1, and the plant height of Leaf-M2 was 6.5% higher than that of MZ-1. At harvest time on July 8th, Leaf-M2 and the control group were very similar in plant height, with Leaf-M2 only 1.1% taller than MZ-1. On the other hand, the plant height of Leaf-M1 increased to approximately 45.0% higher than that of MZ-1, an extremely significant difference from MZ-1 and Leaf-M2. This finding illustrates that the plant height of Leaf-M1 develops slowly in the seedling stage, then increases quickly and significantly exceeds the plant height of MZ-1 at harvest time. The plant height of Leaf-M2 has the slowest rate of development during the seedling stage, but its growth rate following the seedling stage approaches that of MZ-1, reaching approximate equivalence.

**3.4 Analysis of physiological and biochemical traits among the three groups**

SOD (superoxide dismutase) is ubiquitous in the bodies of animals and plants. It can react with free radicals produced in the process of metabolism and is an important antioxidase in living tissues. SOD activity analysis results for the three groups at the late stage of seedling formation are shown in Table 1. SOD activities for Leaf-M1 and Leaf-M2 were significantly lower (about 16.6% and 30.1%) than that of the control group. Specifically, there were significant differences between Leaf-M1 and the control group, and extremely significant differences between Leaf-M2 and the control group. In addition, SOD activity of Leaf-M2 was lower than that of Leaf-M1 and there was a significant difference between the two.

During aging of plant organs or under adverse conditions, membrane lipid peroxidation induced by accumulation of reactive oxygen species can occur. MDA is one of the most important products of membrane lipid peroxidation. MDA content analysis results for the three groups at the late stage of seedling formation are listed in Table 1. The MDA content of Leaf-M1 was about 10.2% lower than that of the control group, a significant difference. The MDA content for Leaf-M2 was about 6.5% higher than that of the control group, also achieving significance. Leaf-M2 showed the highest MDA content, and Leaf-M1 showed the lowest MDA content. The differences in MDA content between Leaf-M1 and Leaf-M2 were significant.

Sugar is both the product of photosynthesis and the substrate of respiration, and it plays an important role in plant growth, development, and metabolism. Reducing sugar has important biological significance because it is the primary energy substance and intermediate metabolite of plants. Results of analysis of reducing sugar content are listed in Table 1. The reducing sugar content of the control group was the highest, while the reducing sugar content of Leaf-M1 and Leaf-M2 was 7.4% and 4.9% lower, respectively. The lowest reducing sugar content was found in the leaves of Leaf-M2. According to difference significance analysis, there were no significant differences among any two of the three groups.

Chlorophyll is an important physiological activator in higher plants and other photosynthetic organisms. Results of the analysis of relative chlorophyll content for the three groups during the late stage of seedling formation are presented in Table 1. Leaf-M2 showed the highest chlorophyll relative content, followed by the control group. Leaf-M1 had the lowest relative chlorophyll content. The relative chlorophyll content of Leaf-M2 was 19.1% higher than that of the control group, while the chlorophyll relative content of Leaf-M1 was 11.3% lower. According to difference significance analysis, there were extremely significant differences between all pairs within the three groups. The high relative chlorophyll content of Leaf-M2 is mainly attributed to changes in surface color.

Table 1 Physiological and biochemical traits of two mutants and control group

| Groups | SOD activity (U/g) | MDA content (μmol/g) | Reducing sugar content (mg/g) | Chlorophyll relative content |
|---|---|---|---|---|
| MZ-1(CK) | 664±16$^{aA}$ | 78.6±3.5$^a$ | 12.2±2.3$^a$ | 33.5±0.4$^A$ |
| Leaf-M1 | 554±39$^{bB}$ | 70.6±7.8$^b$ | 11.3±0.5$^a$ | 29.7±0.2$^B$ |
| Leaf-M2 | 464±43$^{cB}$ | 83.7±2.3$^c$ | 11.6±1.3$^a$ | 39.9±1.2$^C$ |

Note: Different capital letters (B, C) at the superscript indicate extremely significant differences ($P < 0.01$), and different lowercase letters (b, c) at the superscript indicate significant differences ($P < 0.05$), mean ± SD.

### 3.5 Comparison of biomass among three groups at harvest

Processing results of various biomass data for the three groups at harvest time are listed in Table 2. Yield per plant of Leaf-M1 was about 50.9% higher than that of the control group, a significant difference between the two groups. The yield per plant of Leaf-M2 was about 7.8% higher than that of control group, showing no significant difference. Leaf-M2 had an approximately equivalent effective pod number per plant to the control group, whereas the effective pod number per plant for Leaf-M1 was about 43.2% and 45.2% higher than those of the control group and Leaf-M2, respectively, demonstrating extremely significant differences. In terms of yield per pod, the three groups yielded similar means, with no significant differences. Leaf-M2 had the largest pod number per branch, 1.8% higher than that of the control group. The pod number per branch of Leaf-M1 was the lowest, 4.8% lower than that of the control group. The three groups did not vary significantly in terms of mean pod number per branch. Together, these results indicate that the yield per plant of

Leaf-M1 was significantly higher than that of control group, which can be attributed to the greater number of pods per plant.

Table 2 Biomass data of two mutants and MZ-1 at harvest time

| Groups | Effective pod number per plant | Yield per pod (g) | Pod number per branch | Yield per plant (g) |
| --- | --- | --- | --- | --- |
| MZ-1(CK) | 14.8±2.1$^{bB}$ | 0.78±0.08$^b$ | 5.43±0.80$^b$ | 11.6±2.6$^b$ |
| Leaf-M1 | 21.2±4.5$^{cC}$ | 0.83±0.06$^b$ | 5.17±0.35$^b$ | 17.5±3.6$^c$ |
| Leaf-M2 | 14.6±2.5$^{bB}$ | 0.84±0.11$^b$ | 5.53±1.11$^b$ | 12.5±3.6$^b$ |

Note: Different capital letters (B, C) at the superscript indicate extremely significant differences ($P < 0.01$), and different lowercase letters (b, c) at the superscript indicate significant differences ($P < 0.05$), mean ± SD.

### 3.6 Comparison of seed composition among three groups

Nutrient and chemical components of seeds are important factors that must be considered in high-quality breeding. The seed compositions of the two mutants (Leaf-M1 and Leaf-M2, gained from field tests of generation $M_3$) and the control group were tested; results are shown in Table 3. The three groups showed no significant differences in terms of water content and three major nutrients (proteins, fat, and starch). Iron (Fe) content was similar between Leaf-M1 and Leaf-M2, which were 10.9% lower than that of the control group, a difference lacking in significance. Copper (Cu) and magnesium (Mg) content of Leaf-M1 was the highest, 53.1% and 2.9% higher than those of the control group, respectively. Cu and Mg content of Leaf-M2 was 13% and 2.6% lower than those of the control group. In terms of Cu and Mg content in seeds, Leaf-M1 and Leaf-M2 presented no significant differences from the control group, but there were significant differences between Leaf-M1 and Leaf-M2. All three groups had approximately consistent content in terms of phosphorus (P), potassium (K), Zinc (Zn), Manganese (Mn), and calcium (Ca), with no significant differences detected.

Table 3 Seed composition of two mutants and control group in the harvest stage

| Groups | Water (g/100 g) | Protein (g/100 g) | Crude fat (%) | Crude starch (g/100 g) | P (g/100 g) | K (g/100 g) |
| --- | --- | --- | --- | --- | --- | --- |
| MZ-1(CK) | 8.45±0.53$^b$ | 16.70±0.85$^b$ | 1.40±0.21$^b$ | 52.6±1.13$^b$ | 0.332±0.018$^b$ | 1.01±0.24$^b$ |
| Leaf-M1 | 8.83±0.69$^b$ | 16.67±0.14$^b$ | 1.14±0.01$^b$ | 51.7±0.49$^b$ | 0.349±0.018$^b$ | 1.00±0.12$^b$ |
| Leaf-M2 | 8.77±0.39$^b$ | 16.13±0.01$^b$ | 1.30±0.01$^b$ | 52.6±1.41$^b$ | 0.341±0.023$^b$ | 1.04±0.10$^b$ |

Continued Table 3 Seed composition of two mutants and control group in the harvest stage

| Groups | Fe (mg/100 g) | Zn (mg/100 g) | Mn (mg/100 g) | Cu (mg/100 g) | Ca (mg/100 g) | Mg (mg/100 g) |
| --- | --- | --- | --- | --- | --- | --- |
| MZ-1(CK) | 4.97±0.02$^b$ | 3.79±0.74$^b$ | 0.770±0.105$^b$ | 0.877±0.071$^{bc}$ | 38.9±2.2$^b$ | 75.7±0.92$^{bc}$ |
| Leaf-M1 | 4.43±0.14$^c$ | 3.53±0.21$^b$ | 0.773±0.028$^b$ | 1.343±0.495$^b$ | 42.5±2.3$^b$ | 76.9±0.57$^b$ |
| Leaf-M2 | 4.43±0.42$^c$ | 3.23±0.25$^b$ | 0.737±0.136$^b$ | 0.763±0.085$^c$ | 36.6±3.3$^b$ | 73.5±0.11$^c$ |

Note: Different lowercase letters (b, c) at the superscript indicate significant differences ($P < 0.05$), mean ± SD.

## 4. Conclusion

In the field of radiation mutation breeding, few studies have comprehensively investigated the important agronomic traits and the physiological and biochemical traits of mutants throughout the entire life cycle; the relationship between traits and yields also lacks significant study. To meet the needs of breeding studies and other biological studies, a field experiment of the $M_3$ generation of two mutants derived from early neutron radiation of pea seeds was carried out. The results indicated that as compared to the control group, the Leaf-M1 mutant developed significant changes in growth and other physiological and biochemical traits. In particular, the yield per plant of Leaf-M1 was significantly higher than that of the control group. Analytical results indicated that this increase in yield per plant must be related to the increase in pod number per plant, and the increase in pod number per plant was attributed to the increase of branch number per plant. Moreover, the chlorophyll content in the leaves of Leaf-M1 was significantly lower than that of the control group; however, this did not significantly impact its yield. This could be explained by the mutation of the leaves of Leaf-M1 from needle-shaped leaves to broad leaves, with a corresponding increase of leaf area. Similar morphological mutation was not observed in Leaf-M2, but the color of its leaves changed. The change was primarily caused by the increased chlorophyll content in the leaves, which was the primary reason for the increase in yield per plant.

This study proved the breeding value of Leaf-M1 and Leaf-M2. Leaf-M1 was superior to Leaf-M2 in terms of yield, while Leaf-M2 had certain landscape values. Moreover, the two mutants could be used as materials for molecular biological studies, such as those exploring the functional genes of plants. In addition, differences in morphological characteristics between the two mutants and the control group as well as between Leaf-M1 and Leaf-M2 must be related to their physiological characteristics. Further studies on such relationships are still needed in the future.

**Acknowledgements**

This work is Supported by the National Natural Science Foundation of China (No. 11675069) and the Fundamental Research Funds for the Central Universities of China (No. lzujbky-2019-kb09).

**References**

[1] The State of Food Security and Nutrition in the World (2019). http://www.fao.org/home/digital-reports/en/

[2] L. J. Stadler, Mutations in barley induced by X-rays and radium. science **68**, 186-187 (1928). https://doi.org/10.1126/science.68.1756.186

[3] Y. Kazama, T. Hirano, H. Saito, Y. Liu, S. Ohbu, Y. Hayashi, T. Abe, Characterization of highly efficient heavy-ion mutagenesis in Arabidopsis thaliana. BMC Plant Biology **11**, 161 (2011). https://doi.org/10.1186/1471–2229-11-161

[4] J. Si, H. Zhang, Z. Wu, Applications and research progress of plant breeding with ion implantation technique. Journal of Radiation Research and Radiation Processing **30**(6), 321–327 (2012). https://doi.org/10.11889/j.1000-3436.2012.rrj.30.120601

[5] J. S. Wang, L. X. Qiao, L. S. Zhao, P. Wang, B. T. Guo, L. X. Liu, J. M. Sui, Performance of peanut mutants and their offspring generated from mixed high-energy particle field radiation and tissue culture. Genetics and Molecular Research **14**(3), 10837–10848 (2015). https://doi.org/10.4238/2015.September.9.22

[6] L. Yu, W. Li, Y. Du, G. Chen, S. Luo, R. Liu, H. Feng, L. Zhou, Flower color mutants induced by carbon ion beam irradiation of geranium (Pelargonium 3 hortorum, Bailey). Nuclear Science and Techniques **27**(5), 112 (2016). https://doi.org/10.1007/s41365-016-0117-3

[7] X. Bian, A. Tian, J. Geng, L. Ding, J. Gong, J. Hu, T. Cao, The $M_2$ Morphological Variation and RAPD Analysis of Maize Irradiated by Proton. Journal of Anhui Agricultural Sciences **45**(33), 149-152 (2017). https://doi.org/10.3969/j.issn.0517-6611.2017.33.050

[8] Y. Zhang, L. Zhou, S. Wang, B.Mao, Study on Selection of a New Strawberry Strain of Benihoppe Developed by $^{60}$Co-γ Irradiation Induced Mutation Breeding. Journal of Nuclear Agricultural Sciences **32**(8), 1457–1465 (2018). https://doi.org/10.11869/j.issn.100-8551.2018.08.1457

[9] W. Gu, L. Zhou, R. Liu, W. Jin, Y. Qu, X. Dong, W. Li, Synergistic responses of NHX, AKT1, and SOS1 in the control of $Na^+$ homeostasis in sweet sorghum mutants induced by $^{12}C^{6+}$-ion irradiation. Nuclear Science and Techniques **29**, 10 (2018). https://doi.org/10.1007/s41365-017-0341-5

[10] W. Yuan, Y. Wang, Y. Xi, L. Sun, Y. Tu, Human respiratory tract model of uranium-containing compounds and calculation of pulmonary retention under single inhalation. Nuclear Techniques **42**(11), 110302 (2019). https://doi.org/10.11889/j.0253-3219.2019.hjs.42.110302

[11] F. Wang, J. Yang, Q. Xie, S. Zhai, Z. Yang, Analysis and optimization of production conditions for $^{18}$F production using a medical cyclotron. Nuclear Techniques **41**(2), 20301 (2018). https://doi.org/10.11889/j.0253-3219.2018.hjs.41.020301

[12] FAO/IAEA Mutant Variety Database. http//mvgs.iaea.org/

[13] Z. Liu, Y. Yang, L. Zheng, R. Liu, C. Yang, M. Wang, Measurement and analysis of the $^{232}$Th(n,2n) reaction rate in a polyethylene shell with DT neutrons. Nuclear Techniques **41**(6), 60502 (2018). https://doi.org/10.11889/j.0253-3219.2018.hjs.41.060502

[14] W. Zhang, L. Jiao, M. Hoshi, Relative Biological Effectiveness of Induced Micronuclei in Root-tip Cells of Onion Seedlings Irradiated with 0.8 Mev Neutrons. Radiation Protection 26(3), 162-165 (2006). https://doi.org/10.3321/j.issn:1000-8187.2006.03.006

[15] D. Xu, Z. Yao, H. Feng, Y. Yin, Effects of different dosages of neutron radiation on seed germination and seedling growth of needle leaf pea. Chinese Agricultural Science Bulletin 31(12), 200-204 (2015). https://doi.org/10.11924/j.issn.1000-6850.casb14100118

[16] A. Marshak, W. S. Malloch, The Effect of Fast Neutrons on Chromosomes in Meiosis and Its Bearing upon Pachytene Pairing. Genetics **27**(6), 576-583 (1942). https://doi.org/10.1007/BF02982833

[17] S. O. Lochlainn, R. G. Fray, J. P. Hammond, G. J. King, P. J. White, S. D. Young, M. R. Broadley, Generation of nonvernal-obligate, faster-cycling Noccaea caerulescens lines through fast neutron mutagenesis. New Phytologist **189**(2), 409–414 (2011). https://doi.org/10.2307/40983842

[18] Y. Bolon, W. J. Haun, W. W. Xu, D. Grant, M. G. Stacey, R. T. Nelson, D. J. Gerhardt, J. A. Jeddeloh, G. Stacey, G. J. Muehlbauer, J. H. Orf, S. L. Naeve, R. M. Stupar, C. P. Vance, Phenotypic and Genomic Analyses of a Fast Neutron Mutant Population Resource in Soybean. Plant Physiology **156**(1), 240–253 (2011). https://doi.org/10.1104/pp.110.170811

[19] E. J. Belfield, X. Gan, A. Mithan, C. Brown, C. Jiang, K. Franklin, E. Alvey, A. Wibowo, M. Jung, K. Bailey, S. Kalwan, J. Ragoussis, R. Mott, N. P. Harberd, Genome-wide analysis of mutations in mutant lineages selected following fast-neutron irradiation mutagenesis of Arabidopsis thaliana. Genome Research **22**(7), 1306–1315 (2012). https://doi.org/10.1101/gr.131474.111

[20] J. Wang, J. Sui, Y. Xie, H. Guo, L. Qiao, L. Zhao, S. Yu, L. Liu, Generation of peanut mutants by fast neutron irradiation combined with in vitro culture. Journal of Radiation Research **56**(3), 437–445 (2015). https://doi.org/10.1093/jrr/rru121

[21] D. Xu, Z. Yao, Y. Yin, H. Feng, Study on $M_1$ and $M_2$ generation effect of different dosages of neutron radiation on flax seed. Nuclear Techniques **40**(2), 020203 (2017). https://doi.org/10.11889/j.0253–3219.2017.hjs.40.020203

[22] Y. Chen, X. Wang, S. Lu, H. Wang, S. Li, R. Chen, An Array-based Comparative Genomic Hybridization Platform for Efficient Detection of Copy Number Variations in Fast Neutron-induced Medicago truncatula Mutants. Jove-Journal of Visualized Experiments **129**, e56470 (2017). https://doi.org/10.3791/56470

[23] Z. Liu, D. Xu, X. Lian, W. Zhou, S. Kou, Z. Yao, Effects of fast neutron irradiation on seed germination and seedling growth of maize. Journal of Radiation Research and Radiation Processing **36**(2), 020401 (2018). https://doi.org/10.11889/j.1000-3436.2018.rrj.36.020401

[24] X. Yang, R. Ren, Progress of Pea Production and Breeding in oVersea and Inland. Gansu Agricultural Science and Technology (8), 3-5 (2005). https://doi.org/10.3969/j.issn.1001–1463.2005.08.001

[25] P. Gepts, W. D. Beavis, E. C. Brummer, R. C. Shoemaker, H. T.Stalker, N. F. Weeden, N. D. Young, Legumes as a model plant family: genomics for food and feed Report of the Crosslegume Advances through Genomics (ACTG) Conference. Plant Physiology **137**(4), 1228–1235 (2005). https://doi.org/10.1104/pp.105.060871

[26] Y. Qu, Y. Wang, H. Feng, J. Cheng, W. Li, L. An, Effects of UV-B radiation on stems elongation and cell wall polysaccharides of pea seedlings. Journal of Radiation Research and Radiation Processing 30(5), 303–308 (2012). https://doi.org/10.11889/j.1000-3436.2012.rrj.30.120509

[27] M. Bourgault, J. Brandb, M. Tauszc, G. J. Fitzgeraldb, Yield, growth and grain nitrogen response to elevated $CO_2$ of five field pea (*Pisum sativum* L.) cultivars in a low rainfall environment. Field Crops Research **196**, 1–9 (2016). https://doi.org/10.1016/j.fcr.2016.04.011

[28] S. N. Innes, L. E. Arve, B. Zimmermann, L. Nybakken, T. I. Melby, K. A. Solhaug, J. E. Olsena, S. Torre, Elevated air humidity increases UV mediated leaf and DNA damage in pea (*Pisum sativum*) due to reduced flavonoid content and antioxidant power. Photochemical & Photobiological Sciences **18**(4), 387–399 (2019). https://doi.org/10.1039/c8pp00401c

[29] Y. Jiang, D. L. Lindsay, A. R. Davis, Z. Wang, D. E. MacLean, T. D. Warkentin, R. A. Bueckert, Impact of heat stress on pod-based yield components in field pea (*Pisum sativum* L.). Journal of Agronomy and Crop Science **206**(1), 76–89 (2019). https://doi.org/10.1111/jac.12365

[30] P. Wang, P. Sun, Evaluation of radiation effects on peripheral organs of $^{252}$Cf cervical carcinoma under close cavity radiotherapy. Nuclear Techniques **41**(5), 50301 (2018). https://doi.org/10.11889/j.0253-3219.2018.hjs.41.050301

[31] L. Zhang, Z. Wang, S. Xu, J. Yang, C. Lian, Study on Drought-resistance of A New Pea Cultivar with Needle Leaves MZ-1. Gansu Agricultural Science and Technology (12), 13-16 (2009). https://doi.org/10.3969/j.issn.1001-1463.2009.12.005

[32] J. Yin, H. Lu, Q. Xie, J. Ding, N. Li, A Study on Rapid Colorimetric Determination of Water Soluble Total Sugar, Reducing Sugar and Starch in Tobacco with 3,5-dinitrosalicylic Acid. Journal of Yunnan Agricultural University **22**(6), 829–838 (2007). https://doi.org/10.3969/j.issn.1004-390X.2007.06.011